\documentclass[aps,prl,preprintnumbers,groupedaddress,showpacs]{revtex4}
\usepackage{graphicx}
\usepackage{dcolumn}
\usepackage{bm}
\usepackage[latin1]{inputenc}
\usepackage[spanish,english]{babel}
\begin{document}
\def\coup{K}
\def\heli{\Upsilon}
\def\be{\begin{eqnarray}}
\def\ee{\end{eqnarray}}
\preprint{none}

\title{Transiciones de fase topológicas en cristales líquidos}

\author{Ricardo Paredes V.$^1$, Ana I. Fariñas-Sánchez$^1$ and Bertrand Berche$^2$}

\affiliation{$^1$Centro de F\'{\i}sica, Instituto Venezolano de Investigaciones Cient{\'\i}ficas,
Apartado 21827, Caracas 1020A, Venezuela,}

\affiliation{$^2$Laboratorie de Physique des Matériaux, UMR CNRS 7556, Université Henri Poincaré Nancy 1, B. P. 239, F-54506 Vand{\oe}uvre les Nancy cedex, France}

\begin{abstract}
Mediante simulaciones de Monte Carlo, utilizando escalamiento de tamaño finito y transformaciones conforme se reportan evidencias de una transición de fase topológica en cristales líquidos en dos dimensiones. A altas temperaturas se presenta una fase paramagnética mientras que a bajas temperaturas se encuentra un fase de cuasi-largo-orden (QLRO). Se encuentra que a muy bajas temperaturas el exponente de la función de correlación de la fase QLRO es lineal con la temperatura teniéndose de esta manera un comportamiento típico de ondas de espín. Esto último contradice predicciones de que para sistemas con grupo de simetría global no abeliano las ondas de espín no son relevantes. Adicionalmente se discute que implicaciones tendría la presencia de impurezas en este tipo de transiciones.
\end{abstract}
\maketitle

\section{Introducción}

Mermin y Wagner\cite{Mermin-Wagner} establecieron que para sistemas con grupo de simetría continuo no existe fase ferromagnética o de orden a largo alcance (LRO), a temperatura diferente de cero,  para dimensiones menores o iguales que dos. Sin embargo, sistemas de este tipo pueden presentar un tipo de transición que está determinado por la aparición de defectos topológicos en pares a bajas temperaturas que justo en la transición, $T_{KT}$,  se desligan\cite{Berezinskii,Kost-Thou,Kosterlitz}. A este fenómeno se le llama transición de fase topológica o transición de Berezinskii, Kosterlitz y Thouless (BKT). 

El modelo XY, con grupo de simetría $O(2)$, presenta este tipo transición en dos dimensiones\cite{Kosterlitz}. Entre sus características resaltantes  tenemos que la fase   LRO es sustituida, a bajas temperaturas, por un orden a  cuasi-largo-alcance (QLRO). A muy bajas temperaturas, las correlaciones son dominadas por las ondas de espín obteniéndose una dependencia del exponente de la función de correlación, $\eta$, con la temperatura, $T$, de la forma: $\eta = T/2\pi$, donde la constante de Boltzmann $k_B=1$  y  el factor de acoplamiento $J=1$. Debido a este comportamiento QLRO, a temperaturas menores que  $T_{KT}$, la susceptibilidad, que mide las fluctuaciones de la magnetización, diverge. 
Otra característica de este tipo de transición es que a temperaturas justo por encima de la temperatura de transición $t=(T-T_{KT})/T_{KT}\gtrsim 0$, la longitud de correlación, $\xi$, diverge de una manera mucho más fuerte que la típica ley de potencias, $\xi \sim t^{-\nu}$, encontrada en las transiciones de segundo orden. Dicha divergencia es del tipo de  una singularidad escencial: $\xi \sim \exp({b t^{-1/2}})$.

Por otro lado, otro modelo de simetría continua como el de Heisenberg ferromagnético, con grupo de simetría $O(3)$, no presenta ningún tipo de transición de fase en $d=2$ mientras que si presenta la típica transición para-ferromagnética en $d=3$\cite{Polyakov,Wolff2}. Por esta razón surgió la pregunta de que si sistemas con grupo de simetría no abelianos podían presentar transiciones del tipo BKT. Se ha reportado que el modelo de Heisenberg antiferromagnético completamente frustrado presenta una transición del tipo BKT pero sin correlaciones del tipo de ondas de espín a bajas temperaturas\cite{kawamura, wintel}. 

Kunz y Zumbach\cite{KunzZumbach} realizaron un estudio intensivo en $d=2$,  mediante simulaciones,  del modelo $RP^2$ el cual consiste en un sistema con grupo de simetría global $O(3)$ pero con grupo de simetría local $Z_2$. El modelo $RP^2$ describe la transición nem\'atica isotrópica de cistales líquidos en $d=3$. Ellos determinaron la longitud de correlación para $t\gtrsim 0$ y encontraron un buen ajuste para una singularidad escencial. Por otro lado, mediante escalamiento de tamaño finito, estimaron que también era válido un ajuste del tipo ley de potencia. Sin  embargo, en base a cálculos de energía y calor específico así como de ciertas cantidades que estiman el número de defectos topológicos, argumentaron que la transición de fase debería ser del tipo BKT.

En el 2003 se retom\'o el estudio pero ahora utilizando el modelo de Lebwohl-Lasher(LL)\cite{LL} para describir a los cristales l{\'\i}quidos en $d=2$. Este modelo ha sido muy exitoso para detectar la transici\'on de fase discont{\'\i}nua d\'ebil que se observa en los experimentos de cristales l{\'\i}quidos en $d=3$.  En este modelo se representan las mol\'eculas mediante vectores unitarios $\vec{\sigma}_w$ colocados en los sitios de una red hiperc\'ubica $\Lambda$ de longitud $L$. Para este sistema el Hamiltoniano viene dado por:
\begin{equation}
-\frac{H}{k_BT}=\frac{J}{k_BT}\sum\limits_w\sum\limits_\mu P_2(\vec{\sigma}_w\cdot\vec{\sigma}_{w+\mu}),
\end{equation}
donde $P_2$ es el segundo polinomio de Legendre y la interacci\'on es de primeros vecinos. En $d=2$, se determinó\cite{PLA} la presencia de un orden del tipo QLRO a bajas temperaturas con soluciones del tipo de ondas de espín cuando $T$ tiende a cero. Esto se hizo mediante el novedoso método de las transformaciones conformes. De esta manera se concluy\'o, que sistemas con grupo de simetría no abeliano tenían transición BKT con correlaciones del tipo de ondas de espín a bajas temperaturas.
 
En el presente artículo revisará el método de transformaciones conformes y su utilidad para determinar exponentes de la función correlación para sistemas con invarianza de escalas. Luego estudiando el parámetro de orden nemático bajo $T_{KT}$, mediante escalamiento de tamaño finito, se obtendrá de nuevo el exponente de la correlación y se mostrará la excelente concordancia entre los dos procedimientos. Finalmente se discutirá el efecto de la presencia de impurezas sobre este tipo de transición. 

\section{Método de las transformaciones conformes}
Uno de los grandes problemas que existe al simular un sistema f{\'i}sico es el de tener evidencia de lo que ocurre en el l{\'\i}mite termodin\'amico. Por lo general, en los estudios de fen\'omenos cr{\'\i}ticos, se simulan sistemas a diferentes tama\~nos y se estudian como escalan las cantidades termodin\'amicas en funci\'on de $L$. El exponente de escalamiento de dichas cantidades tiene relaci\'on con los exponentes cr{\'\i}ticos asociados a dichas cantidades. El costo computacional es demasiado alto. Recientemente, para sistemas en $d=2$ se ha comenzado a utilizar las transformaciones conformes (TC). Est\'a t\'ecnica consiste en realizar simulaciones en sistemas finitos y conectar los resultados con los de un sistema infinito v{\'\i}a una TC. Esto lo podemos realizar en sistemas que presenten invarianza de escalas y para $d=2$. 

Sistemas con invarianza de escala cumplen con la hip\'otesis de homogeneidad para la funci\'on de correlaci\'on de dos puntos para cualquier densidad $\phi$, tal como el par\'ametro de orden, la energ{\'\i}a, etc., $\langle \phi(b\vec{r}_2)\phi(b\vec{r}_1)\rangle=b^{-\eta}\langle\phi(\vec{r}_2)\phi(\vec{r}_1)\rangle$, donde $b$ es un factor de escala y $\eta$ es la dimensi\'on an\'omala o exponente de la correlaci\'on. De la misma manera, si existe invarianza de escala,  podemos relacionar mediante la hip\'otesis de homogeneidad la funci\'on de correlaci\'on entre sistemas que se conectan bajo una transformaci\'on conforme. Supongamos que las simulaciones se realizan en un sistema cuadrado ($w=u+iv$) de tama\~no $L\times L$ ($-L/2 \le u \le L/2, ~0 \le v \le L$) y mediante una transformaci\'on de Schwarz-Christoffel lo mapeamos en el plano semi-infinito ($z=x+iy, ~ 0 \le y <  \infty$). Para esta transformaci\'on el cambio de escala es local y las funciones de correlaci\'on se relacionan  mediante $\langle\phi(w_1)\phi(w_2)\rangle=|w'(z_1)|^{-x_\phi}|w'(z_2)|^{-x_\phi}\langle\phi(z_1)\phi(z_2)\rangle$, donde $x_\phi=\frac{1}{2}\eta$.

Si las simulaciones se realizan en una red cuadrada con condiciones de borde fijas se obtiene que el perfil de  densidad (funci\'on de correlaci\'on de un punto)  se comporta como una ley de potencia de la forma: $\langle\phi(w)\rangle_{\scriptsize \mbox{sq}} \sim \kappa(w)^{-\eta/2}$ con $\kappa(w)=\Im[z(w)](|1-z(w)^2||1-k^2z(w)^2|)^{-1/2}$, donde $z(w)=\mbox{sn}(2Kw/L)$, siendo $K$ la integral el{\'\i}ptica completa de primera clase, sn el seno el{\'\i}ptico de Jacobi y $k$ una constante. Esta expresi\'on  se obtiene  ya que el perfil para una red semi-infinita con condiciones de borde fijas se conoce exactamente $\langle \phi(z)\rangle_{\frac{1}{2}\infty}\sim y^{-x_\phi}$\cite{Cardy}.

\begin{figure*}[h]
\includegraphics[width=7cm]{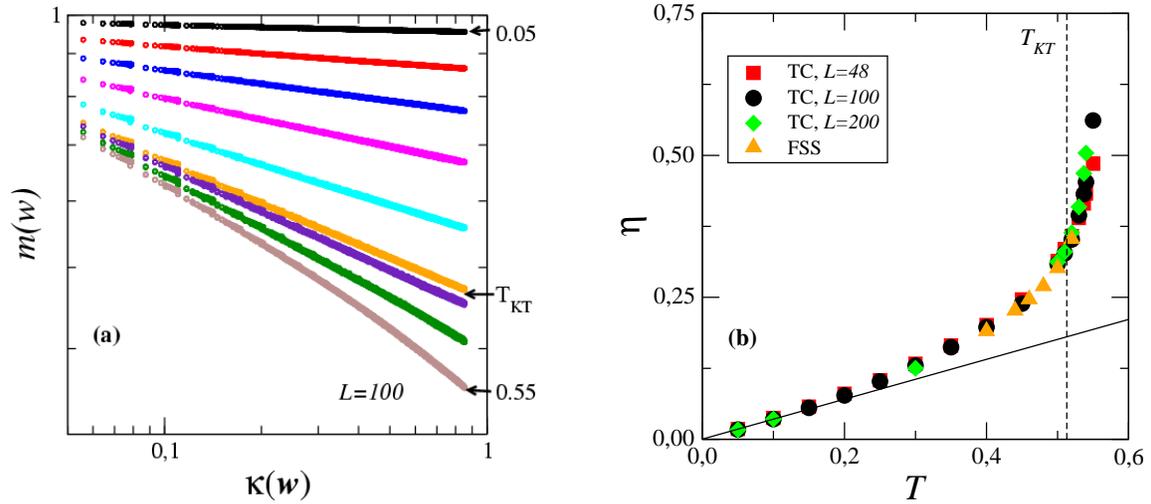}\hspace{10mm}
\includegraphics[width=7cm]{Fig1b.eps}
\caption{a) Par\'ametro de orden $m(w)$ local en funci\'on de $\kappa(w)$ para diferentes temperaturas. Se nota un comportamiento del tipo ley de potencias para valores de temperatura menores que el valor $T_{KT}=0.513$ obtenido por Kunz y Zumbach\cite{KunzZumbach}. b) Exponente de la funci\'on de correlaci\'on como funci\'on de la temperatura utilizando los ajustes de la Fig. 1(b) junto con los obtenidos para $L=48$ y $L=200$. Los triangulos representan los estimados hechos a partir de escalamiento de tama\~no finito. A temperaturas cercanas a cero se obseva el comportamiento $\eta \propto T$.}
\end{figure*}

Gracias a la invarianza de escalas presente en las fases del tipo QLRO, la metodolog{\'\i}a de las TC se aplic\'o con exito en la determinaci\'on de los exponentes de la funci\'on de correlaci\'on del modelo XY en $d=2$ a $t<0$\cite{EPL,JPA}. M\'as recientemente, utilizando las TC en el modelo  LL para los cristales l{\'\i}quidos en $d=2$, se concluy\'o la existencia de un orden QLRO en este sistema\cite{PLA}. Para este problema la funci\'on de correlaci\'on de un punto utilizada es: $m(w)=\langle P_2({\vec \sigma}_w\cdot{\vec h}_{\partial\Lambda(w)})\rangle_{\scriptsize \mbox{sq}}$ donde ${\vec h}_{\partial\Lambda(w)}$ indica que todas las mol\'eculas en el borde $\partial\Lambda(w)$ tienen la orientaci\'on fija ${\vec h}$. En la Fig. 1(a) se muestran los perfiles del par\'ametro de orden como funci\'on de $\kappa$. Se nota, como a temperaturas menores que $T_{KT}$, se presenta  un comportamiento del tipo ley de potencias mientras que muy por encima de esta temperatura dicho comportamiento es abandonado indicando que el sistema se encuentra en una regi\'on donde no existe invarianza de escalas. La determinaci\'on del cambio de comportamiento se puede hacer de manera cuantitativa calculando el $\chi^2$\cite{PLA}. En la Fig. 1(b) se grafican los exponentes de la correlaci\'on obtenidos a partir de los ajustes de la Fig. 1(a) como funci\'on de $T$. A muy bajas temperaturas se observa el comportamiento lineal t{\'\i}pico de las ondas de esp{\'\i}n.

\section{Escalamiento de tama\~no finito}

En el presente trabajo se realizaron simulaciones del modelo LL en dos dimensiones con condiciones de borde peri\'odicas. Las simulaciones se hicieron para temperaturas comprendidas entre $T=0.4$ y $T=0.58$. Se simularon sistemas con longitudes comprendidas entre $L=16$ y $L=512$. Se empleo el algoritmo de Wolff\cite{Wolff}. Se tomaron del orden de $10^5$ pasos de equilibraci\'on y $10^6$ pasos de promediaci\'on. Los tiempos de relajaci\'on para todos los tama\~nos y longitudes no sobrepasaron los $200$ pasos de Monte Carlo. En particular, se reportan estimados  del exponente de la funci\'on de correlaci\'on del par\'ametro de orden utilizando la t\'ecnica de escalamiento de tama\~no finito para el par\'ametro de orden.

Un buen par\'ametro de orden para describir la transici\'on de fase nem\'atica-isotr\'opica, en $d=3$, para el modelo LL es:
\begin{equation}
M=L^{-2}\left\langle \sum\limits_w P_2({\vec\sigma}_w\cdot{\hat n})\right\rangle = L^{-2}\left\langle \sum\limits_w P_2(\cos\theta_w)\right\rangle,
\end{equation}
donde ${\hat n}$ es un vector unitario que indica la direcci\'on preferencial y se le denomina director y $\theta_w$ es el \'angulo entre ${\vec \sigma}_w$ y el director. $M$ tiende a $1$ a muy bajas temperaturas y a $0$ a altas temperaturas. En $d=2$, $M$ deber{\'\i}a ser $0$ para todo $T$ ya que no puede haber rompimiento de la simetr{\'\i}a continua\cite{Mermin-Wagner}. 

En la Fig 2.(a) se tiene una gr\'afica del par\'ametro como funci\'on de la temperatura. Se observa un valor finito del par\'ametro de orden a temperaturas bajas. Sin embargo, la tendencia es que disminuya al aumentar $L$. En la Fig. 2(b) se muestra el comportamiento del par\'ametro de orden con la longitud del sistema. Obs\'ervese que para temperaturas bajas el comportamiento es del tipo ley de potencia. Esto indica la presencia de invarianza de escalas para $T<T_{KT}$ o lo que es lo mismo un orden del tipo QLRO. Sobre $T_{KT}$ la caida es exponencial, comportamiento caracter{\'\i}stico de una fase paramagn\'etica. El exponente de la ley de potencia obtenido a bajas temperauras es el correspondiente a la funci\'on de correlaci\'on de un punto, es decir, $\frac{1}{2}\eta$. Colocando los exponentes provenientes de los ajustes realizados con los datos de la Fig. 2(b) en la Fig. 1(b) tenemos que la correspondencia entre las dos metodolog{\'\i}as reportadas  es excelente.

\begin{figure*}[h]
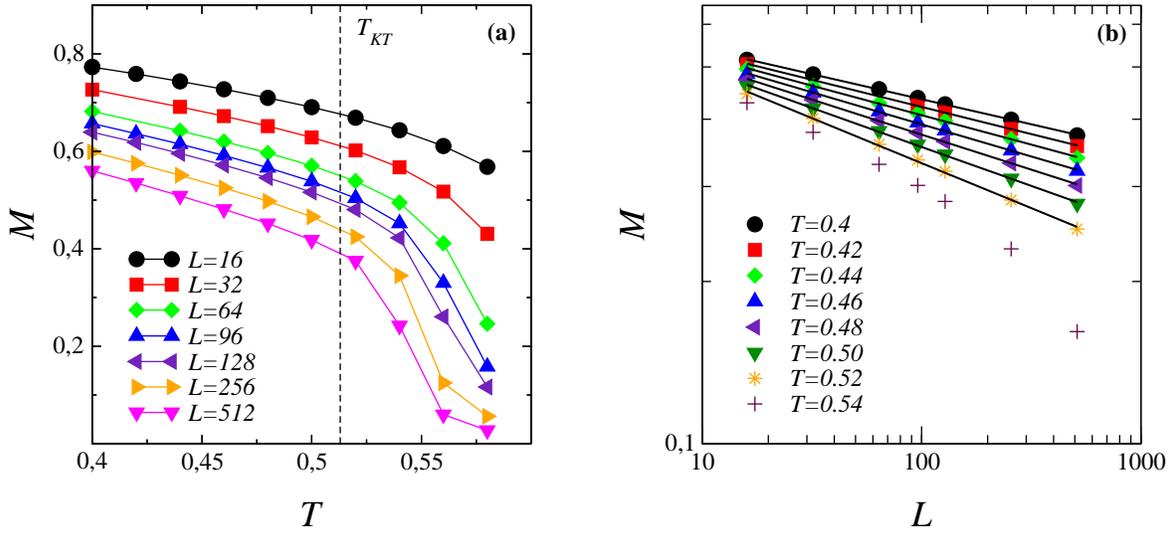

\includegraphics[width=7cm]{Fig2a.eps}\hspace{10mm}
\includegraphics[width=7.3cm]{Fig2b.eps}
\caption{a) Par\'ametro de orden en funci\'on de la temperatura para diferentes tama\~no $L$ del sistema. Se observa como al incrementar $L$ el valor de $M$ para cada $T$ disminuye. El valor de $T_{KT}$ mostrado es el estimado por Kunz y Zumbach \cite{KunzZumbach}. b) Para cada $T$ el par\'ametro de orden en funci\'on de $L$. Comportamiento del tipo ley de potencia para $T<T_{KT}$. Por encima  de la temperatura cr{\'\i}tica la tendencia es la de un comportamiento exponencial.}
\end{figure*}
\section{Discusi\'on}

En el presente trabajo se reportan evidencias de que a bajas temperaturas existe un orden del tipo QLRO. La manera m\'as eficiente de realizar estos estimados es mediante el uso de la t\'ecnica de las transformaciones conformes ya que solo se necesita hacer simulaciones en un solo tama\~no de red, inclusive para los primeros resultados bast\'o el uso de una red peque\~na ($L=48$). El costo computacional del m\'etodo de escalamiento de tama\~no finito es mucho mayor al tener que simular para muchas longitudes de red y tama\~nos mayores. Realizando estudios del escalamiento de la susceptibilidad se tiene un porcedimiento alternativo para el c\'alculo del exponente $\eta$ al igual de una forma muy precisa para determinar que en $T_{KT}$ la singularidad es escencial\cite{PRB}. Por lo tanto, definitivamente se concluye que, para los cristales l{\'\i}quidos en $d=2$, se tiene un sistema con grupo de simetr{\'\i}a global no-abeliano que presenta una transici\'on BKT con orden QLRO a bajas temperaturas.

En la literatura se ha reportado que para sistemas con transic\'on de tipo BKT, la introducci\'on de desorden en los enlaces es totalmente irrelevante\cite{EPJB}. Esto se debe a que debido a la singularidad escencial para $t>0$ el exponente del calor espec{\'\i}fico $\alpha\rightarrow -\infty$ ($2-d\nu=\alpha$) y debido al criterio de Harris\cite{harris} los exponentes cr{\'\i}ticos del sistema puro no deber{\'\i}an  cambiar. Sin embargo, debido a cambios que aparecen en la coordinaci\'on el valor de $T_{KT}$ deber{\'\i}a depender de la intensidad, $c$, del desorden. Si se realizaran simulaciones del LL con enlaces aleatorios obtendr{\'\i}amos que el exponente de la correlaci\'on ser{\'\i}a id\'entico para todos los valores $T_{KT}(c)$.

Donde resulta muy interesante discutir el problema del desorden ser{\'\i}a en $d=3$ con campo magn\'etico aleatorio. Seg\'un el criterio de Imry y Ma \cite{ImryMa} la fase nem\'atica deber{\'\i}a desaparecer y podr{\'\i}a ser sustituida por un orden del tipo QLRO\cite{chakrabarti,feldman}.

\section{Agradecimientos}
Este trabajo cuenta con el apoyo del programa PCP Venezolano-Franc\'es titulado ``Fluidos Petroleros''.

\end{document}